\documentclass[lettersize,journal]{IEEEtran}
\usepackage{amsthm}
\usepackage{amsmath,amssymb,amsfonts}
\usepackage[linesnumbered,ruled,vlined]{algorithm2e}
\usepackage{array}
\usepackage{xcolor}
\usepackage{textcomp}
\usepackage{subfig}
\usepackage{stfloats}
\usepackage{url}
\usepackage{verbatim}
\usepackage{graphicx}
\usepackage{makecell}
\usepackage{multirow}
\usepackage{cite}
\usepackage{booktabs}
\usepackage{diagbox}
\newcommand{\figsubref}[2]{Fig.~\ref{#1}\subref{#2}}

\newtheorem{proposition}{Proposition}

\usepackage{etoolbox}
\makeatletter
\def\IEEEtitletopspace{-0.1in}

\newcommand{\patchwarning}[1]{\typeout{*** PATCH WARNING: #1 ***}}

\patchcmd{\@maketitle}
  {\vskip 1.0em\par}
  {\vskip 0.1em\par}
  {}{\patchwarning{Pattern '\string\vskip 1.0em\string\par' not found in \string\@maketitle}}

\patchcmd{\@maketitle}
  {\par \addvspace {0.5\baselineskip }}
  {\par \addvspace {-1.6\baselineskip }}
  {}{\patchwarning{Pattern '\string\par \string\addvspace\space\{0.2\string\baselineskip\space\}' not found in \string\@maketitle}}
\makeatother
\begin{document}
\setlength{\textfloatsep}{8pt}
\setlength{\floatsep}{6pt}
\setlength{\intextsep}{8pt}

\title{Learning-based Multiuser Beamforming for Holographic MIMO~Systems}

\author{Shiyong~Chen,~\IEEEmembership{Student Member,~IEEE,}
        and Shengqian~Han,~\IEEEmembership{Senior Member,~IEEE}%
\thanks{The authors are with the School of Electronics and Information Engineering, Beihang University, Beijing 100191, China (Email: \{shiyongchen, sqhan\}@buaa.edu.cn).}%
%\thanks{Manuscript received April 19, 2025; revised August 16, 2025.}
}

%Shell \MakeLowercase{\textit{et al.}}: A Sample Article Using IEEEtran.cls for IEEE Journals}

% Remember, if you use this, you must call \IEEEpubidadjcol in the second

 \maketitle
 \vspace{-3cm}
\begin{abstract}
Holographic multiple-input multiple-output (HMIMO) can improve spectral efficiency (SE) with low hardware cost, but conventional alternating optimization (AO) methods for jointly optimizing digital and holographic beamformers are computationally expensive. Learning-based beamforming offers a low-complexity alternative, and graph neural networks (GNNs) are particularly attractive because they can exploit permutation equivariance (PE). The optimal HMIMO beamforming policy exhibits PE properties across multiple dimensions. Existing methods either use high-dimensional GNNs, increasing model size and training complexity, or exploit only partial PE properties, leading to performance degradation. To address this issue, we reformulate the problem by learning an equivalent beamformer that removes the RF-chain dimension from the network output while preserving the PE property of the original problem. The reformulation introduces a nontrivial column-space constraint because the equivalent beamformer must be representable by the phase-pattern matrix. We then develop a cascaded architecture consisting of a gradient-based graph neural network (GGNN) and two projection modules. The GGNN jointly learns the holographic and equivalent beamformers using update equations motivated by their coupled gradient structures, while the projection modules recover the digital beamformer and enforce the column-space and transmit-power constraints. Simulation results show that the proposed method achieves higher SE with lower inference latency than the AO baseline and exhibits better generalization than existing learning-based baselines.
\end{abstract}
\vspace{-0.2cm}

\vspace{-0.3cm}
\section{Introduction}
\IEEEPARstart{M}{assive} multiple-input multiple-output (mMIMO) is an effective technique for improving system spectral efficiency (SE)~\cite{An_overview}. However, as the number of antennas increases, conventional mMIMO architectures face growing challenges in power consumption, hardware cost, and physical size~\cite{Hybrid_precoder}. To address these issues, holographic MIMO (HMIMO) has emerged as a promising alternative. An HMIMO surface employs densely integrated subwavelength antenna elements controlled by a simple diode-based circuit for efficient wave manipulation. This hardware structure can significantly reduce implementation complexity and power consumption~\cite{Holographic_MIMO_Communications}. The channel characteristics and capacity limits of HMIMO were studied in~\cite{Holographic_MIMO_Communications_With}, demonstrating the SE potential of densely integrated surfaces.

To exploit the potential of HMIMO, prior work has developed efficient beamforming methods~\cite{Three_controller,Uplink,Amplitude_control_RHS, Decomposition, Wideband}. In~\cite{Three_controller}, the antenna elements are modeled as polarizable dipoles with Lorentzian polarizability, enabling beamforming design under Lorentzian-phase, amplitude-only, or binary-amplitude constraints. This model provides a theoretical foundation for subsequent algorithm design. In~\cite{Uplink}, an uplink multi-user HMIMO system was considered, where an alternating optimization (AO) approach jointly optimized the digital combiner and holographic beamformer to maximize the sum rate. In~\cite{Amplitude_control_RHS} and~\cite{Decomposition}, the digital and holographic beamformers were alternately optimized in a downlink multi-user system. Furthermore, AO was applied to a HMIMO-assisted multi-user wideband system in~\cite{Wideband}, where the analog and baseband combiners on each subcarrier were alternately optimized. Nevertheless, the iterative process in such AO methods results in high computational complexity, hindering real-time implementation.

Learning-based beamforming offers a low-latency alternative to iterative AO by replacing repeated optimization with a fixed-depth inference pass. Graph neural networks (GNNs) are particularly suitable for this purpose because they can exploit permutation equivariance (PE) and permutation invariance (PI), which commonly arise in beamforming policies~\cite{Understanding_the_04,MDGNN}. In HMIMO systems, the optimal beamforming policy exhibits PE properties across multiple dimensions. Existing PE-based designs mainly follow two routes. One route uses high-dimensional input and output representations to capture multi-dimensional PE properties~\cite{MDGNN}, but this increases model size and training complexity. The other route simplifies the network by exploiting PE only along partial dimensions~\cite{Graph_neural_03}, which does not fully match the PE properties of the optimal HMIMO beamforming policy.

To resolve this performance-complexity tradeoff, we first reformulate the beamforming
optimization problem by combining the phase pattern (i.e., the phase distribution across the holographic surface) and the digital beamformer into a single equivalent beamformer. This reformulation removes the RF-chain dimension from the network output, but it also imposes a column-space constraint on the equivalent beamformer. To handle this constraint together with the transmit-power constraint, we propose a network architecture composed of a gradient-based GNN (GGNN) followed by two cascaded projection modules. Different from existing gradient-inspired GNNs that typically update a single beamforming variable~\cite{Gradient_GNN}, the proposed GGNN is tailored to the coupled updates of the holographic and equivalent beamformers. The two projection modules then enforce the column-space constraint and the transmit-power constraint, respectively. Simulation results show that the proposed method achieves higher SE with significantly shorter inference latency than the AO algorithm and exhibits better generalization than other learning-based~baselines.

\vspace{-0.1cm}
\section{System Model and Problem Formulation}  \label{System Model}
Consider a downlink HMIMO system where a base station (BS) equipped with a holographic surface serves multiple single-antenna users.
% as illustrated in Fig.~\ref{hybrid_architecture}.
The HMIMO surface functions as a leaky-wave antenna and consists of three main components: $N_{\mathrm{RF}}$ feeds, a waveguide, and $N_t = N_x N_y$ antenna elements, where $N_x$ and $N_y$ denote the number of elements along the $x$- and $y$-axes, respectively~\cite{Holographic_MIMO_Communications}.

Let $\mathbf{s} = [s_1,\ldots,s_K]^{\mathsf{T}} \in \mathbb{C}^{K\times1}$ denote the modulated symbol vector for $K$ users, where $s_k$ is the symbol intended for user $k$, $\mathbb{E}[
\|s_k\|^2]=1$, and $\|\cdot\|$ denotes the Frobenius norm. After baseband processing, the baseband signal is given by $\mathbf{V}\mathbf{s}$, where $\mathbf{V} = [\mathbf{v}_1, \ldots, \mathbf{v}_K] \in \mathbb{C}^{N_{\mathrm{RF}}\times K}$ is the \textit{digital beamformer} and $\mathbf{v}_k \in \mathbb{C}^{N_{\mathrm{RF}}\times 1}$ denotes the digital beamforming vector for user~$k$. The baseband signal is subsequently upconverted to the carrier frequency through $N_{\mathrm{RF}}$ radio-frequency (RF) chains. Each feed, connected to a dedicated RF chain, then converts the high-frequency signals into reference waves. The waveguide supports in-plane propagation of these reference waves, and the antenna elements equipped with diode-based controllers regulate the waves leaked into free space, thereby enabling dynamic holographic beamforming~\cite{Decomposition}. 

The resulting beam pattern is determined by both the amplitude and the phase at each element. The phase distribution over the holographic surface, referred to as the phase pattern, is denoted by the constant matrix $\mathbf{M}_{\mathrm{p}}\in\mathbb{C}^{N_t\times N_{\mathrm{RF}}}$, whose $(i,l)$-th entry is given by $[\mathbf{M}_{\mathrm{p}}]_{i,l}=e^{-j\mathbf{k}_s\cdot \mathbf{r}^{\,l}_{m,n}}$, where $\quad l\in\{1,\ldots,N_{\mathrm{RF}}\}$, $i=(m-1)N_y+n$ indexes the $(m,n)$-th radiation element with $m\in\{1,\ldots,N_x\}$ and $n\in\{1,\ldots,N_y\}$, and $e^{-j\mathbf{k}_s\cdot \mathbf{r}^{\,l}_{m,n}}$ denotes the phase of the reference wave emitted from feed $l$ and observed at element $(m,n)$, with $\mathbf{k}_s$ and $\mathbf{r}^{\,l}_{m,n}$ representing the wave vector and the position vector from feed $l$ to element $(m,n)$, respectively. The amplitude of each element is controlled by a real-valued \textit{holographic beamformer} $\mathbf{a}=\left[a_{1}, \ldots, a_{N_t}\right]^{\mathsf{T}}$, where the $((m-1)N_y+n)$-th entry denotes the amplitude of the $(m,n)$-th radiation element. 

The received signal at the users is expressed~as~\cite{Amplitude_control_RHS} 
\begin{equation} \label{received signal}
    \mathbf{y} = \mathbf{H}^{\mathsf{H}}\mathrm{diag}(\mathbf{a})\mathbf{M}_{\mathrm{p}}\mathbf{V}\mathbf{s}+\mathbf{n},
\end{equation}
where $\mathbf{H}=\left[ \mathbf{h}_1,\ldots,\mathbf{h}_K \right] \in \mathbb{C} ^{N_t\times K}$ denotes the channel matrix, $\mathbf{h}_k \in \mathbb{C} ^{N_t\times 1}$ is the channel vector between the BS and user $k$, $\mathbf{n} \sim \mathcal{CN}(0, \sigma_n^2\mathbf{I}_{K})$ is the white noise with variance $\sigma_n^2$, and $\mathbf{I}_{K}$ is the identity matrix of size $K\times K$.

The channel between the BS and user $k$ is modeled as~\cite{Amplitude_control_RHS}
\begin{equation} \label{channel}
    \mathbf{h}_k = \sqrt{N_t/{I}}\sum\nolimits_{i=1}^{I} \alpha^k_{i}\, \mathbf{b}(\theta^k_{\text{AoD},i}, \phi^k_{\text{AoD},i}),
\end{equation}
where $I$ is the number of multipath components, \(\alpha^k_{i}\) is the complex channel gain of the $i$-th path, and $\theta^k_{\text{AoD}, i}$ and $\phi^k_{\text{AoD}, i}$ denote azimuth and elevation angles of departure (AoD) from the BS to the user $k$, respectively. The transmit steering vector $\mathbf{b}(\theta,\phi)$ is defined as~\cite{Amplitude_control_RHS}
\begin{equation}
\mathbf{b}(\theta, \phi) = \sqrt{1/N_t} \left[e^{jk_f d_{1,1}}, \ldots, e^{jk_f d_{N_x,N_y}}\right]^{\mathsf{T}},
\end{equation}
with $d_{m,n}=(m-1)d_x\sin \theta\cos\phi+(n-1)d_y\sin \theta\sin \phi$, where $d_x$ and $d_y$ denote the element spacing along the $x$- and $y$-axes, respectively, and $k_f$ is the free-space wavenumber.

%\subsection{Problem Formulation}
With~\eqref{received signal}, the joint optimization problem for the digital beamformer $\mathbf{V}$ and the holographic beamformer $\mathbf{a}$, aiming to maximize the SE, can be formulated as~\cite{Amplitude_control_RHS}
\begin{subequations} \label{P1}
    \begin{align}
\!\max_{\mathbf{a},\mathbf{V}}\ \!\!&\sum\limits_{k=1}^K \!\log_2\!\bigg(\!\!1\!+\!\frac{\left\| \mathbf{h}_{k}^{\mathsf{H}}\mathrm{diag}\left( \mathbf{a} \right) \mathbf{M}_{\mathrm{p}}\mathbf{v}_k \right\|^2}{\sum_{j=1,j\ne k}^{K}\!{\left\| \mathbf{h}_{k}^{\mathsf{H}}\mathrm{diag}\left( \mathbf{a} \right) \mathbf{M}_{\mathrm{p}}\mathbf{v}_j \right\|^2}\!\!+\!\sigma_n^2} \!\bigg)
  \label{P1:object}\\
    \mathrm{s}.\mathrm{t}.\
    &0\leq a_{i}\leq 1, \quad i\in\{1,\ldots,N_t\}, \label{P1:ampli}\\
    &\left\|\mathrm{diag}\left( \mathbf{a} \right) \mathbf{M}_{\mathrm{p}} \mathbf{V} \right\|^{2} \leq P_{\max},\label{P1:power}
    \end{align}
\end{subequations}
where $P_{\max}$ denotes the maximum transmit power of the BS. 

The optimal beamforming policy for problem~\eqref{P1} is defined~as
\begin{equation}\label{optimal_policy}
(\mathbf{a}^{\star},\mathbf{V}^{\star})=\mathcal{F}(\mathbf{H},\mathbf{M}_{\mathrm{p}}),
\end{equation}
where $\mathbf{a}^{\star}$ and $\mathbf{V}^{\star}$ are the optimal solutions corresponding to the input $(\mathbf{H},\mathbf{M}_{\mathrm{p}})$.

\section{Learning the Beamforming Policy with GNNs}
In this section, we begin with analyzing the PE properties of the optimal beamforming policy. To learn a policy that satisfies these properties while maintaining low training complexity, we transform problem~\eqref{P1} into an equivalent form and propose a novel neural network architecture.

%\vspace{-0.15cm}
\subsection{Properties of the Optimal Beamforming Policy} \label{GNN limitation}
The PE property of a policy is defined such that if the indices of its input are permuted, then permuting the indices of the output accordingly yields corresponding optimal~output.

The following proposition indicates that the policy~\eqref{optimal_policy} exhibits a three-dimensional PE (3DPE)~property, expressed~as
\begin{equation}\label{3D PE}
\big(\boldsymbol{\Pi}_{N_t}^{\mathsf{T}}\mathbf{a}^{\star},\boldsymbol{\Pi}_{\mathrm{RF}}^{\mathsf{T}}\mathbf{V}^{\star}\boldsymbol{\Pi}_K\big)\!=\!\mathcal{F}\!\big(\boldsymbol{\Pi}_{N_t}^{\mathsf{T}}\mathbf{H}\boldsymbol{\Pi}_K,\boldsymbol{\Pi}_{N_t}^{\mathsf{T}}\mathbf{M}_{\mathrm{p}}\boldsymbol{\Pi}_{\mathrm{RF}}\big),
\end{equation}
where $\boldsymbol{\Pi}_{K}\in\mathbb{R}^{K\times K}$, $\boldsymbol{\Pi}_{N_t}\in\mathbb{R}^{N_t\times N_t}$, and $\boldsymbol{\Pi}_{\mathrm{RF}}\in\mathbb{R}^{N_{\mathrm{RF}}\times N_{\mathrm{RF}}}$ are permutation matrices for users, antennas, and RF chains, respectively. Each permutation matrix is binary with exactly one ``1'' in every row and column, so multiplying by it only reorders the corresponding dimension.

\begin{proposition}\label{proposition1}
When the inputs of the beamforming policy are permuted as $\hat{\mathbf{H}}=\boldsymbol{\Pi}_{N_t}^{\mathsf{T}}\mathbf{H}\boldsymbol{\Pi}_{K}$, $\hat{\mathbf{M}}_{\mathrm{p}}=\boldsymbol{\Pi}_{N_t}^{\mathsf{T}}\mathbf{M}_{\mathrm{p}}\boldsymbol{\Pi}_{\mathrm{RF}}$, then the beamformers $\hat{\mathbf{a}}^{\star}=\boldsymbol{\Pi}_{N_t}^{\mathsf{T}}\mathbf{a}^{\star}$ and $\hat{\mathbf{V}}^{\star}=\boldsymbol{\Pi}^{\mathsf{T}}_{\mathrm{RF}}\mathbf{V}^{\star}\boldsymbol{\Pi}_{K}$ are the optimal solutions of problem~\eqref{P1}.
\end{proposition}
\begin{IEEEproof}
The proof follows from the invariance of the objective and constraints under the corresponding index permutations and is omitted for brevity.
\end{IEEEproof}
A direct way to exploit the 3DPE property is to adopt the multi-dimensional GNN in~\cite{MDGNN}. However, this approach relies on multi-dimensional input and output representations, leading to a large model and high training complexity.

%\vspace{-0.15cm}
\subsection{Problem Transformation}
To fully exploit the PE properties while avoiding increases in model size and training complexity, we transform problem~\eqref{P1} to reduce the output dimensionality of the beamforming policy by combining the phase pattern $\mathbf{M}_{\mathrm{p}}$ and the digital beamformer $\mathbf{V}$ into a single equivalent beamformer, denoted as $\mathbf{V}_{\mathrm{e}}$. Then, problem~\eqref{P1} can be transformed as
\begin{subequations} \label{P2}
    \begin{align}
\max_{\mathbf{a},\mathbf{V}_{{\mathrm{e}}}}\ &\!\sum\limits_{k=1}^K \log_2\!\!\bigg(\!\!1+\!\frac{\left\|\mathbf{h}_{k}^{\mathsf{H}}\mathrm{diag}\left( \mathbf{a} \right) \mathbf v_{\mathrm e,k} \right\|^2}{\sum_{j=1,j\ne k}^{K}\!{\left\| \mathbf{h}_{k}^{\mathsf{H}}\mathrm{diag}\left( \mathbf{a} \right) \mathbf v_{\mathrm e,j} \right\|^2}\!\!+\!\sigma_n^2}\!\bigg)
  \label{P2:object}\\
    \mathrm{s}.\mathrm{t}.\
    &0\leq a_{i}\leq 1, \quad i\in\{1,\ldots,N_t\},  \label{P2:ampli}\\    &\mathbf V_{\mathrm e}=\mathbf{M}_{\mathrm{p}}\mathbf{V},\label{P2:range}\\ 
    &\left\|\mathrm{diag}\left( \mathbf{a} \right) \mathbf{V}_{\mathrm{e}} \right\|^{2} \leq P_{\max}, \label{P2:power}
    \end{align}
\end{subequations}
where $\mathbf{V}_{\mathrm{e}} = [\mathbf{v}_{\mathrm{e},1}, \ldots, \mathbf{v}_{\mathrm{e},K}] \in \mathbb{C}^{N_t \times K}$, $\mathbf{v}_{\mathrm{e},k}\in \mathbb{C}^{N_t \times 1}$ denotes the equivalent beamforming vector for user $k$, and~\eqref{P2:range} constrains $\mathbf V_{\mathrm e}$ to lie in the column space of $\mathbf M_{\mathrm p}$.

The optimal beamforming policy for problem~\eqref{P2} is defined as~$(\mathbf{a}^{\star},\mathbf{V}^{\star}_\mathrm{e})=\mathcal{F}_{\mathrm{e}}(\mathbf{H},\mathbf{M}_{\mathrm{p}})$, where $\mathbf{a}^{\star}$ and $\mathbf{V}^{\star}_\mathrm{e}$ represent the optimal solutions. Compared with the policy $\mathcal{F}(\cdot,\cdot)$ in~\eqref{optimal_policy} for problem~\eqref{P1} where the outputs involve user, antenna, and RF-chain dimensions, the outputs of the new policy $\mathcal{F}_{\mathrm{e}}(\cdot,\cdot)$ include only the user and antenna dimensions, with the RF-chain dimension eliminated. 

%\vspace{-0.3cm}
\subsection{Overall Network Architecture}
\begin{figure}%[htbp]
\centering
\includegraphics[width=0.45\textwidth]{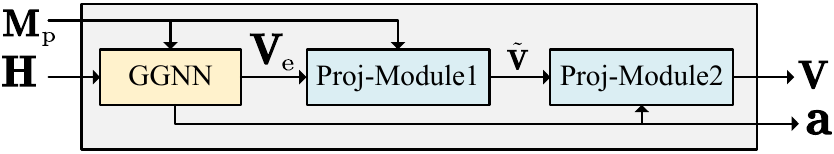}
\caption{Overall network architecture.}  \label{MB_DNN}
% \vspace{-0.5cm}
\end{figure}

To learn $\mathbf{a}$ and $\mathbf{V}_{\mathrm{e}}$ and then recover $\mathbf{V}$ from $\mathbf{V}_{\mathrm{e}}$, we propose a novel network architecture, as illustrated in Fig.~\ref{MB_DNN}. 

The first component is the proposed GNN, called GGNN (to be detailed in the next subsection), which is designed to learn the beamforming policy $\mathcal{F}_{\mathrm{e}}(\cdot,\cdot)$ and output $\mathbf{a}$ and $\mathbf{V}_{\mathrm{e}}$. The constraint~\eqref{P2:ampli} for $\mathbf{a}$ is satisfied via a bounded activation function. The constraint in~\eqref{P2:range} requires $\mathbf V_{\mathrm e}$ to lie in the column space of $\mathbf M_{\mathrm p}$. To satisfy this, we introduce a projection module, which is cascaded after GGNN, to solve the following projection problem~\cite{Uplink}
\begin{equation}\label{LS}
\min_{\mathbf{V}} \left\| \mathbf{V}_{\mathrm{e}}-\mathbf{M}_{\mathrm{p}}\mathbf{V}  \right\|^2.
\end{equation}
The closed-form solution to problem~\eqref{LS}, i.e., the output of the projection module, can be obtained as $\tilde{\mathbf{V}} = (\mathbf{M}_{\mathrm{p}}^{\mathsf{H}} \mathbf{M}_{\mathrm{p}})^{-1} \mathbf{M}_{\mathrm{p}}^{\mathsf{H}} \mathbf{V}_{\mathrm{e}}$, where $\mathbf{M}_{\mathrm{p}}$ is of full column rank since~$N_t\gg N_{\mathrm{RF}}$. 

The resulting $\tilde{\mathbf{V}}$ may not satisfy the power constraint in~\eqref{P1:power}. Therefore, we employ a second module to normalize $\tilde{\mathbf{V}}$ as $ \mathbf{V}= \frac{\tilde{\mathbf{V}} \sqrt{P_{\max}}}{\left\| \mathrm{diag}  (\mathbf{a})\mathbf{M}_{\mathrm{p}}\tilde{\mathbf{V}} \right\|}$.

Next, we show that the overall network architecture satisfies the 3DPE property in~\eqref{3D PE}.
First, as detailed later, the proposed GGNN satisfies the PE property of the beamforming policy $\mathcal{F}_{\mathrm{e}}(\cdot,\cdot)$. Specifically, permutations applied to the antenna and user dimensions of the GGNN inputs induce the same permutations at its output, whereas permutations of the RF-chain dimension in $\mathbf{M}_{\mathrm{p}}$ do not affect the output. Such a joint PE and PI (PEPI) property can be expressed as
\begin{equation}\label{Ve PE}
\!\!\big(\boldsymbol{\Pi}_{N_t}^{\mathsf{T}}\mathbf{a}^{\star}, \boldsymbol{\Pi}_{N_t}^{\mathsf{T}}\mathbf{V}^{\star}_{\mathrm{e}}\boldsymbol{\Pi}_K \big)\!\!=\!\mathcal{F}_{\mathrm{e}}\big(\boldsymbol{\Pi}_{N_t}^{\mathsf{T}}\mathbf{H}\boldsymbol{\Pi}_K, \boldsymbol{\Pi}_{N_t}^{\mathsf{T}}\mathbf{M}_{\mathrm{p}}\boldsymbol{\Pi}_{\mathrm{RF}}\big)\!.
\end{equation}

Second, by examining the operations of the two projection modules, i.e., the closed-form solution of problem~\eqref{LS} and the beamformer normalization, we find that their inputs and outputs satisfy the following properties,~respectively,
\begin{subequations} \label{PE of projection modules}
\begin{align}
&\big(\boldsymbol{\Pi}_{\mathrm{RF}}^{\mathsf{T}}\tilde{\mathbf{V}}\boldsymbol{\Pi}_K\!\big)=\mathcal{F}_{\mathrm{p}_1}\big(\boldsymbol{\Pi}_{N_t}^{\mathsf{T}}\mathbf{V}_{\mathrm{e}}^{\star}\boldsymbol{\Pi}_K,\boldsymbol{\Pi}_{N_t}^{\mathsf{T}}\mathbf{M}_{\mathrm{p}}\boldsymbol{\Pi}_{\mathrm{RF}}\big),\label{PE of the first projection module}\\
&\!\!\!\big(\boldsymbol{\Pi}_{\mathrm{RF}}^{\mathsf{T}}\mathbf{V}\boldsymbol{\Pi}_K\!\big)\!=\!\mathcal{F}_{\mathrm{p}_2}\!\big(\boldsymbol{\Pi}_{\mathrm{RF}}^{\mathsf{T}}\tilde{\mathbf{V}}\boldsymbol{\Pi}_K,\!\boldsymbol{\Pi}_{N_t}^{\mathsf{T}}\mathbf{a}^{\star},\!\boldsymbol{\Pi}_{N_t}^{\mathsf{T}}\mathbf{M}_{\mathrm{p}}\boldsymbol{\Pi}_{\mathrm{RF}}\!\big).\label{PE of the second projection module}
\end{align}
\end{subequations}

Combining~\eqref{Ve PE} with~\eqref{PE of projection modules} yields the following:
(i) Permuting the antenna dimension of $\mathbf H$ and $\mathbf M_{\mathrm p}$ induces the same permutation on $\mathbf{a}^{\star}$.
(ii) Permuting the user dimension of $\mathbf H$ induces the corresponding permutation on $\mathbf V_{\mathrm e}^{\star}$, which permutes $\tilde{\mathbf V}$ via~\eqref{PE of the first projection module} and then permutes $\mathbf V$ via~\eqref{PE of the second projection module}.
(iii) Permuting the RF-chain dimension of $\mathbf M_{\mathrm p}$ induces the same permutation on $\tilde{\mathbf V}$ in~\eqref{PE of the first projection module} and, together with the permuted $\mathbf M_{\mathrm p}$, yields the same permutation on $\mathbf V$ in~\eqref{PE of the second projection module}. Therefore, the PE properties satisfied by the overall network architecture are consistent with the 3DPE property of the original optimal beamforming policy as shown in~\eqref{3D PE}.

%\vspace{-0.2cm}
\subsection{Design of GGNN} \label{GS-GNN}
\begin{figure}[htbp]
\centering
\includegraphics[width=0.35\textwidth]{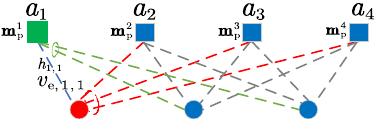}
\caption{Illustration of a graph with $N_t=4$ and $K=3$.}  \label{GNN_graph}
% \vspace{-0.5cm}
\end{figure}

We present the design of the proposed GGNN in this subsection. As shown in Fig.~\ref{GNN_graph}, the GGNN operates on a graph with two types of vertices, namely antenna vertices and user vertices, connected by edges. Features and actions are associated with both antenna vertices and edges. Specifically, for each edge $(n,k)$ connecting antenna vertex $n$ and user vertex $k$, the feature is the channel coefficient $h_{n,k}$, namely the $n$-th entry of $\mathbf{h}_k$, and the corresponding action is the beamforming coefficient $v_{\mathrm{e},n,k}$, namely the $n$-th entry of $\mathbf{v}_{\mathrm{e},k}$. For antenna vertex $n$, the feature is $\mathbf{m}^n_{\mathrm{p}}\in \mathbb{C}^{1\times N_{\mathrm{RF}}}$ denoting the $n$-th row of $\mathbf{M}_{\mathrm{p}}$, and the action is $a_n$. User vertices are associated with neither features nor actions.

For comparison, we first introduce a conventional GNN that learns $\mathbf{V}_{\mathrm{e}}$ and $\mathbf{a}$ on the same graph. Let $\overline{\mathbf{v}}_{\mathrm{e},i,k}^{l}\in\mathbb{C}^{C_l\times1}$ and $\overline{\mathbf{a}}_{n}^{l}\in\mathbb{C}^{C_l\times1}$ denote the hidden representations of edge $(i,k)$ and antenna vertex $n$ at layer $l$, respectively. Their update equations are given by
\begin{subequations}\label{update_equation_of_GNN}
    \begin{align}
\overline{\mathbf{v}}_{\mathrm{e},n,k}^{l+1}
\!&\textstyle=\!\sigma\Big(\mathbf{S}^l\overline{\mathbf{v}}_{\mathrm{e},n,k}^{l}
\!+\!\mathbf{P}_1^l\!\!\!\!\!\!\underset{\text{antenna aggregation}}{\underbrace{\sum\limits_{i=1,\,i\neq n}^{N_t}\!\!\!\!\!\overline{\mathbf{v}}_{\mathrm{e},i,k}^{l}}}\!+\!\mathbf{P}_2^l\!\!\!\!\!\underset{\text{user aggregation}}{\underbrace{\sum\limits_{j=1,\,j\neq k}^{K}\!\!\!\!\!\overline{\mathbf{v}}_{\mathrm{e},n,j}^{l}}}\Big), \label{edge update equation}\\
\overline{\mathbf{a}}_{n}^{l+1}&\textstyle=\sigma\Big(\mathbf{W}_{1}^{l}\overline{\mathbf{a}}_{n}^{l}
+\mathbf{W}_{2}^{l}
\underset{\text{vertex aggregation}}{\underbrace{\sum\nolimits_{k=1}^{K}\overline{\mathbf{v}}_{\mathrm{e},n,k}^{l}}}\Big), \label{vertex update equation}
    \end{align}
\end{subequations}
where $C_l$ denotes the feature dimension of the $l$-th layer, $\mathbf{S}^{l}, \mathbf{P}_{m}^{l}, \mathbf{W}_m^{l}\in\mathbb{C}^{C_{l+1}\times C_l}, m\in\{1,2\}$ are trainable matrices, and $\sigma(\cdot)$ is the activation function. For complex-valued hidden representations, $\sigma(\cdot)$ is applied separately to the real and imaginary parts. The input features $\overline{\mathbf{v}}_{\mathrm{e},n,k}^{0}$ and $\overline{\mathbf{a}}_{n}^{0}$ are $h_{n,k}$ and $\mathbf{m}^n_{\mathrm{p}}$, respectively. In an $L$-layer GNN, the output representation $\overline{\mathbf{v}}_{\mathrm{e},n,k}^{L}$ corresponds to $v_{\mathrm{e},n,k}$, while $a_n$ is obtained by applying a bounded activation function to $\|\overline{\mathbf{a}}_{n}^{L}\|$, ensuring~$0\leq a_i\leq 1,i\in\{1, \ldots, N_t\}$.

As analyzed in~\cite{Gradient_GNN}, the update equation of GNNs and the gradient descent iteration equation have similarities, which can be exploited for GNN design. Building on this insight, we derive the gradient descent iteration equations for $\mathbf{V}_{\mathrm{e}}$ and $\mathbf{a}$, and use them to design the information aggregation mechanism in the GGNN. Specifically, for the $k$-th user, the gradient descent iteration equation with respect to $\mathbf{V}_{\mathrm{e}}$ is given~by
\begin{equation}
\mathbf{v}_{\mathrm{e},k}^{l+1}=\mathbf{v}_{\mathrm{e},k}^{l} + \alpha\nabla_{\mathbf{v}_{\mathrm{e},k}^{l}}f(\mathbf{V}_{\mathrm{e}}),
\label{Gradient iteration}
\end{equation}	
where $\mathbf{v}_{\mathrm{e},k}^{l}
=\bigl[v_{\mathrm{e},1,k}^l, \cdots, v_{\mathrm{e},N_t,k}^l\bigr]^{\mathsf T}$
denotes the equivalent beamformer for user $k$ at the $l$-th iteration,
$\alpha$ is the step size, and $f(\mathbf{V}_{\mathrm{e}})$ denotes the objective function of problem~\eqref{P2}. $\nabla_{\mathbf{v}_{\mathrm{e},k}^{l}} f(\mathbf{V}_{\mathrm{e}})$ is its gradient
with respect to $\mathbf{v}_{\mathrm{e},k}^{l}$, which~is
\begin{equation}\label{Gradient}
\begin{aligned}
& \nabla_{\mathbf{v}_{\mathrm{e},k}^{l}}f(\mathbf{V}_{\mathrm{e}})\!\!=\!\! \beta_{1,k}^l\!\!\left(\tilde{\mathbf{h}}_{k}^{l,\mathsf{H}}\!\!\mathbf{v}_{\mathrm{e},k}^{l}\! \right) \tilde{\mathbf{h}}^{l}_k
\!\!+\!\! \!\!\!\!\sum\limits_{j=1,\!\ j\ne k}^K\!\!\!\!\! \beta_{2,j}^l\!\left( \tilde{\mathbf{h}}_{j}^{l,\mathsf{H}}\mathbf{v}_{\mathrm{e},k}^{l} \right) \tilde{\mathbf{h}}^{l}_j\\
&    =\beta_{1,k}^l\biggl(\sum_{i=1}^{N_t} \tilde{h}^{l,\ast}_{i,k}v_{\mathrm{e},i,k}^{l}\!\biggr) \tilde{\mathbf{h}}^{l}_k\!\!+\!\!\!\!\!\!\sum\limits_{j=1,\!\ j\ne k}^K \!\!\beta_{2,j}^l\left( \tilde{\mathbf{h}}_{j}^{l,\mathsf{H}}\mathbf{v}_{\mathrm{e},k}^{l} \right) \tilde{\mathbf{h}}^{l}_j,
\end{aligned}
\end{equation}
where $\textstyle\beta_{2,j}^l=\frac{1}{ \sum\nolimits_{i=1}^K{\left\| \tilde{\mathbf{h}}_{j}^{l,\mathsf{H}}\mathbf{v}_{\mathrm{e},i}^{l} \right\|^2+\sigma_n^2}}\cdot  \frac{\left\| \tilde{\mathbf{h}}_{j}^{l,\mathsf{H}}\mathbf{v}_{\mathrm{e},j}^{l} \right\|^2}{\sum\nolimits_{i=1,i\ne j}^K{\left\| \tilde{\mathbf{h}}_{j}^{l,\mathsf{H}}\mathbf{v}_{\mathrm{e},i}^{l} \right\|^2+\sigma_n^2}}$ and $\textstyle\beta_{1,k}^l=\frac{1}{ \sum\nolimits_{j=1,j\ne k}^K{\left\| \tilde{\mathbf{h}}_{k}^{l,\mathsf{H}}\mathbf{v}_{\mathrm{e},j}^{l} \right\|^2+\sigma_n^2}}$ are scalar coefficients, $\tilde{\mathbf{h}}_{k}^{l}=\mathrm{diag}\left(\mathbf{a}^{l}\right)\mathbf{h}_{k}$ with $\tilde{h}^l_{i,k}$ being its $i$-th entry, and $\mathbf{a}^{l}\!\!=\!\!\left[a_{1}^l, \ldots, a_{N_t}^l\right]^{\mathsf{T}}$ is the holographic beamformer at iteration $l$.

By substituting \eqref{Gradient} into \eqref{Gradient iteration}, we obtain the explicit gradient descent iteration equation for $\mathbf{v}_{\mathrm{e},k}^{l+1}$. In particular, the $n$-th entry of $\mathbf{v}_{\mathrm{e},k}^{l+1}$ is updated~as
\begin{equation}\label{Full Gradient iteration element0}
\begin{split}
&v_{\mathrm{e},n,k}^{l+1}\!=\!(1\!+\!\alpha\beta_{1,k}^l \|\tilde{h}^l_{n,k}\|^2)v_{\mathrm{e},n,k}^{l}+\!\!\!\!\!\sum\limits_{i=1, i\neq n}^{N_t}\!\!\!\!\big(\alpha\beta_{1,k}^l\tilde{h}^l_{n,k}\cdot\\ 
&v_{\mathrm{e},i,k}^{l}\big)\tilde{h}^{l,\ast}_{i,k}+\!\!\!\!\!\sum_{j=1,\ j\ne k}^K\!\!\!\!\!\alpha\beta_{2,j}^l {\Big( \sum\nolimits_{i=1}^{N_t} (\tilde{h}^l_{i,j})^{\ast}v_{\mathrm{e},i,k}^{l} \Big)\tilde{h}^l_{n,j}}.
\end{split}
\end{equation}

The iteration equation in~\eqref{Full Gradient iteration element0} exhibits a structure analogous to the conventional GNN update in~\eqref{edge update equation}, both involving information aggregation across neighboring indices followed by feature combination. To exploit this analogy, we lift the scalar variables $v_{\mathrm{e},i,k}^{l}$ and $a_{n}^{l}$ in~\eqref{Full Gradient iteration element0} to the hidden representations $\overline{\mathbf{v}}_{\mathrm{e},i,k}^{l}\in\mathbb{C}^{C_l\times1}$ and $\overline{\mathbf{a}}_{n}^{l}\in\mathbb{C}^{C_l\times1}$, respectively, and introduce trainable matrices. The resulting GGNN edge update equation~is\footnote{The proposed GGNN is a finite-layer neural network rather than an iterative optimization algorithm. Hence, inference requires only a fixed-depth forward pass and does not involve a convergence criterion. Although GGNN training is iterative, our simulations show that the training loss converges within a small number of epochs. The corresponding results are omitted due to space limitations.}
\begin{equation}\label{update equation for edges}
\begin{split}\raisetag{22pt}
&\textstyle\overline{\mathbf{v}}_{\mathrm{e},n,k}^{l+1}\!=\!\sigma\Big(\!\mathbf{S}^l\overline{\mathbf{v}}_{\mathrm{e},n,k}^{l}\!+\!\mathbf{P}_1^l\!\!\!\sum\limits_{i=1, i\neq n}^{N_t}\!\!\!\!\mathbf{c}^l_{i,n,k}\!+\!\mathbf{P}_2^l\!\!\sum\limits_{j=1,j\ne k}^K\!\!\!\!\mathbf{d}^l_{j,n,k}\Big),
\end{split}
\end{equation}
where $\mathbf{d}^l_{j,n,k}={\!\big(\sum_{i=1}^{N_t} ({h}^{\ast}_{i,j}\overline{\mathbf{a}}_i^l)\odot\overline{\mathbf{v}}_{\mathrm{e},i,k}^{l}\big)\odot h_{n,j}\overline{\mathbf{a}}_n^l}$, $\mathbf{c}^l_{i,n,k}=h_{n,k}\overline{\mathbf{a}}_n^l\odot\overline{\mathbf{v}}_{\mathrm{e},i,k}^{l}\odot(h^{\ast}_{i,k}\overline{\mathbf{a}}_i^l)$, and $\odot$ denotes the Hadamard product. The step size $\alpha$ and the coefficients $\beta_{1,k}^l$ and $\beta_{2,j}^l$ in~\eqref{Full Gradient iteration element0} are absorbed into the trainable parameter matrices.

The same approach is applied to the holographic beamformer. The gradient descent iteration for $a^l_{n}$ can be written~as
\begin{equation} \label{Analogy Gradient iteration element}
\begin{split}
&a_{n}^{l+1}\!=\!a_{n}^{l}\!-\!\delta\! \sum\nolimits_{k=1}^K\!\!\Big(\!
\mathfrak{R} \big\{
\big( \mathbf{h}_{k}^{\mathsf{T}}\mathrm{diag}\left( \mathbf{a}^l \right)
\mathbf{v}_{\mathrm{e},k}^{l,\ast} \big)
h_{n,k}^{\ast}v_{\mathrm{e},n,k}^{l}
\big\} \\
&-\sum\nolimits_{j=1,j\ne k}^K
\mathfrak{R} \big\{
\big( \mathbf{h}_{k}^{\mathsf{T}}\mathrm{diag}\left( \mathbf{a}^l \right)
\mathbf{v}_{\mathrm{e},k}^{l,\ast} \big)
h_{n,k}^{\ast}v_{\mathrm{e},n,j}^{l}
\big\}\Big),
\end{split}
\end{equation}	
where $\mathfrak{R}\{\cdot\}$ denotes the real-part operator. Comparing~\eqref{Analogy Gradient iteration element} with~\eqref{vertex update equation} suggests the corresponding vertex update for GGNN. By lifting the scalar variables to hidden representations and introducing trainable matrices, the vertex update equation of GGNN is given by
\begin{equation}\label{update equation for vertices}
\begin{split}\raisetag{18pt}
&\textstyle\overline{\mathbf{a}}_{n}^{l+1}=\sigma\big( \mathbf{W}_{1}^{l}\overline{\mathbf{a}}_{n}^{l}+\mathbf{W}_{2}^{l}\sum\nolimits_{k=1}^K\mathbf{f}_{n,k}\big),
\end{split}
\end{equation}
where $\mathbf{f}_{n,k}=\mathfrak{R} \big\{\overline{\mathbf{V}}_{\mathrm{e},k}^{l,\mathsf{H}} \overline{\mathbf{A}}^l\!\odot\!(\mathbf{h}_{k}\mathbf{1}_{C_l}^{\mathsf{T}}) h_{n,k}^{\ast}\!\odot\!\overline{\mathbf{v}}_{\mathrm{e},n,k}^{l} \big\}-\sum\nolimits_{j=1,j\ne k}^K{\mathfrak{R} \big\{\overline{\mathbf{V}}_{\mathrm{e},j}^{l,\mathsf{H}}\overline{\mathbf{A}}^l\odot(\mathbf{h}_{k}\mathbf{1}_{C_l}^{\mathsf{T}})h_{n,k}^{\ast}\odot\overline{\mathbf{v}}_{\mathrm{e},n,j}^{l} \big\}}$, $\mathbf{1}_{C_l}$ denotes the all-ones vector of length $C_l$, $\overline{\mathbf{A}}^{l}=[\overline{\mathbf{a}}_{1}^{l},\cdots, \overline{\mathbf{a}}_{N_t}^{l}]^{\mathsf{T}}\in\mathbb{C}^{N_t\times C_l}$, and $\overline{\mathbf{V}}_{\mathrm{e},k}^{l}=[\overline{\mathbf{v}}_{\mathrm{e},1,k}^{l},\cdots,\overline{\mathbf{v}}_{\mathrm{e},N_t,k}^{l}]^{\mathsf{T}}\in\mathbb{C}^{N_t\times C_l}$.

\subsection{Computational Complexity Analysis}
This subsection analyzes the inference complexity of the proposed architecture in terms of floating-point operations~(FLOPs).

The proposed architecture consists of the GGNN and two projection modules. For the $l$-th GGNN layer, the edge update in~\eqref{update equation for edges} requires $KN_tC_l(K+C_{l+1}+3)$ FLOPs, while the vertex update in~\eqref{update equation for vertices} requires $C_l(2KN_tC_l+5KN_t+3N_tC_{l+1}+K^2)$ FLOPs. Therefore, the total FLOPs of the $L$-layer GGNN are $\sum_{l=1}^{L} C_l(K^2N_t+KN_tC_{l+1}+8KN_t+2KN_tC_l+3N_tC_{l+1}+K^2)$.
For the first projection module, since $\mathbf{M}_{\mathrm{p}}$ is fixed across samples, the matrix
$(\mathbf{M}_{\mathrm{p}}^{\mathsf{H}}\mathbf{M}_{\mathrm{p}})^{-1}\mathbf{M}_{\mathrm{p}}^{\mathsf{H}}$
can be precomputed offline. Thus, the online complexity of this module is reduced to $N_tN_{\mathrm{RF}}K$. For the second projection module, the power normalization requires $N_tN_{\mathrm{RF}}K+2N_tK+N_{\mathrm{RF}}K$ FLOPs.

When $C_l=C_{l+1}=C$, the total number of FLOPs of the proposed architecture is simplified as $LC(K^2N_t+3KN_tC+8KN_t+3N_tC+K^2)
+2N_tN_{\mathrm{RF}}K+2N_tK+N_{\mathrm{RF}}K$.
The resulting Big-O complexity orders of the proposed method and baselines (as described in Sec.~IV-B) are summarized in Table~\ref{Inference_Complexity}. It can be observed that the proposed method 
achieves a much lower inference complexity, scaling linearly with $N_t$. In contrast, the AO method scales cubically with $N_t$. 

\begin{table}[htbp]	
\captionsetup{font=small}
\centering 
\small
\caption{Inference Complexity in Order of Magnitude of FLOPs}	
\begin{tabular}{c|c}
\hline\hline
Name & Inference complexity  \\       
\cline{1-2}
\textbf{Proposed} 
& $\mathcal{O}(LCKN_t(K+C)+N_tN_{\mathrm{RF}}K)$   \\ 
\cline{1-2}
\textbf{VAGNN}    
& $\mathcal{O}(LC^2KN_t+N_tN_{\mathrm{RF}}K)$   \\ 
\cline{1-2}
\textbf{DecGNN}~\cite{Graph_neural_03}   
& \makecell[c]{
$\mathcal{O}(LC^2K+N_tN_{\mathrm{RF}}K)+$\\
$L_{d}(N_tN_{\mathrm{RF}}^2+N_{\mathrm{RF}}^3+N_tN_{\mathrm{RF}}K+N_{\mathrm{RF}}^2K)$
}  \\ 
\cline{1-2}
\textbf{MDGNN}~\cite{MDGNN}    
& $\mathcal{O}(LC^2KN_tN_{\mathrm{RF}})$    \\ 
\cline{1-2}
\textbf{TNN}~\cite{TNN}    
& $\mathcal{O}\!\left(LCN_t(K+N_{\mathrm{RF}})\left(C
+N_tK+N_tN_{\mathrm{RF}}\right)\right)$  \\
\cline{1-2}
\textbf{AO}~\cite{Amplitude_control_RHS}       
& $\mathcal{O}(K^2+N_t^3)$   \\ 
\hline\hline
\end{tabular}
\begin{minipage}{0.95\linewidth}
\quad\footnotesize \textit{Note:} $L_{d}$ denotes the number of iterations required to decompose the learned beamformer into the digital and holographic beamformers.
\end{minipage}
\label{Inference_Complexity}
\end{table}

%\vspace{-0.16cm}
\section{Simulation Results}
In this section, we evaluate the performance of the proposed method and compare it with relevant baselines.

\subsection{Simulation Setup}
Unless otherwise specified, the following simulation setup is used. The HMIMO system is configured with $N_x\!=\!N_y\!=\!12$ antenna elements along the $x$- and $y$-directions, with element spacing $d_x\!=\!d_y\!=\!0.25$\,cm. The numbers of users and RF chains are set to $K\! =\! N_{\mathrm{RF}}\! = \!4$. The reference wave vector satisfies $|\mathbf{k}_s| \!=\! \sqrt{3} k_f\!=\! 2\sqrt{3}\pi f / c$, where the carrier frequency is $f \!=\! 30$ GHz and $c$ denotes the speed of light. The channel model includes two multipath components per user, including one line-of-sight (LoS) path and one non-line-of-sight (NLoS) path. The LoS path gain is independently drawn as $\alpha_i^k \!\sim\! \mathcal{CN}(0, 1)$ and the NLoS path gain is drawn as $\alpha_i^k\!\sim\! \mathcal{CN}(0, 0.01)$~\cite{Amplitude_control_RHS}. The azimuth and elevation AoDs, $\theta^k_{\text{AoD},i}$ and $\phi^k_{\text{AoD},i}$, are drawn from the uniform distribution $\mathcal{U}(-\frac{\pi}{2}, \frac{\pi}{2})$. The signal-to-noise ratio (SNR) is set to~20~dB.

The hyperparameters of the proposed GNN are set as follows. Six hidden layers are used with dimensions $[64, 128, 512, 512, 128, 64]$, respectively. Tanh activation functions are applied to all hidden layers. The GNN is trained in an unsupervised manner, where the loss function is defined as the negative of the objective in~\eqref{P1:object}.\footnote{Unsupervised training avoids the need to generate large labeled datasets with computationally expensive iterative algorithms such as AO. It also
prevents the learned model from being constrained by AO labels, which are not guaranteed to be globally optimal.} A total of 500,000 samples are used for training and another 10,000 for testing. Training is conducted using the Adam optimizer with an initial learning rate of $10^{-3}$ and a batch size of 128.

\subsection{Learning Performance}
 \begin{itemize}
    \item \textbf{VAGNN}: An ablation variant of the proposed architecture that replaces the GGNN with the conventional GNN using the update equations in~\eqref{update_equation_of_GNN}, while retaining the two projection modules. This baseline isolates the gain brought by the gradient-inspired updates.
    
    \item \textbf{DecGNN}: The GNN proposed in~\cite{Graph_neural_03} exploits only the PE along the user dimension to learn a fully digital beamformer, which is then decomposed into digital and holographic beamformers through iterative algorithms.

    \item \textbf{MDGNN}: The multi-dimensional GNN based on the design principle from~\cite{MDGNN}, which satisfies the 3DPE property by extending both the input and output into multi-dimensional representations.

    \item \textbf{TNN}: A transformer-based neural network using the encoder architecture in~\cite{TNN} is adapted as a non-GNN baseline. It directly learns $\mathbf V$ and $\mathbf a$.

    \item \textbf{AO}: The AO algorithm in~\cite{Amplitude_control_RHS}, which iteratively optimizes the digital and holographic~beamformers.
\end{itemize}

\begin{figure}[htbp]\centering
\subfloat[SE versus SNR.]{
\includegraphics[width=0.45\linewidth]{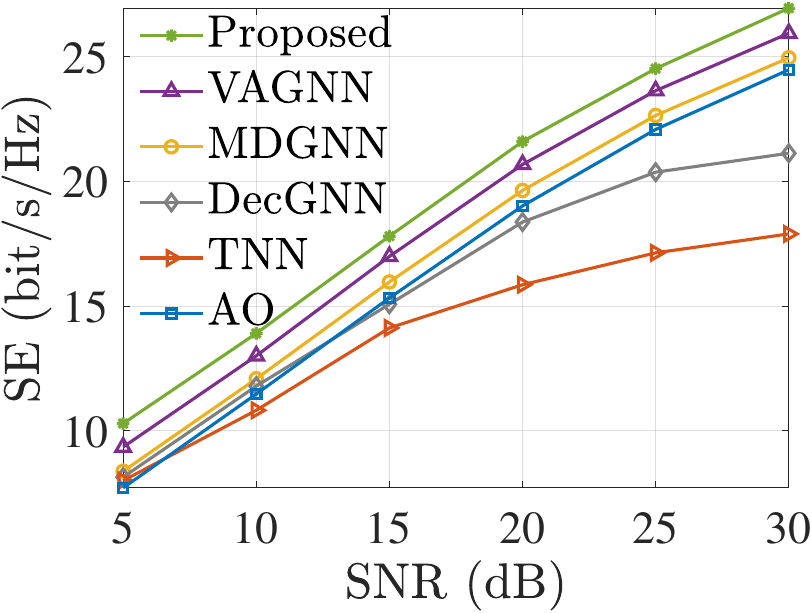}
\label{SE_SNR}
}
\hspace{0.01\linewidth}
\subfloat[SE versus $N_x$ ($N_x=N_y$).]{
\includegraphics[width=0.45\linewidth]{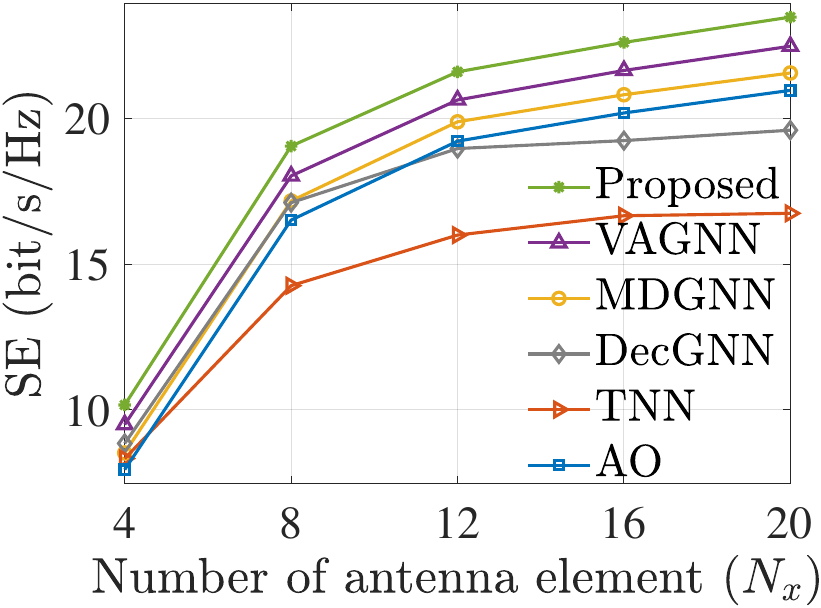}
\label{SE_MN}
}

%\vspace{-0.1cm}

\subfloat[SE versus $N_{\mathrm{RF}}$.]{
\includegraphics[width=0.45\linewidth]{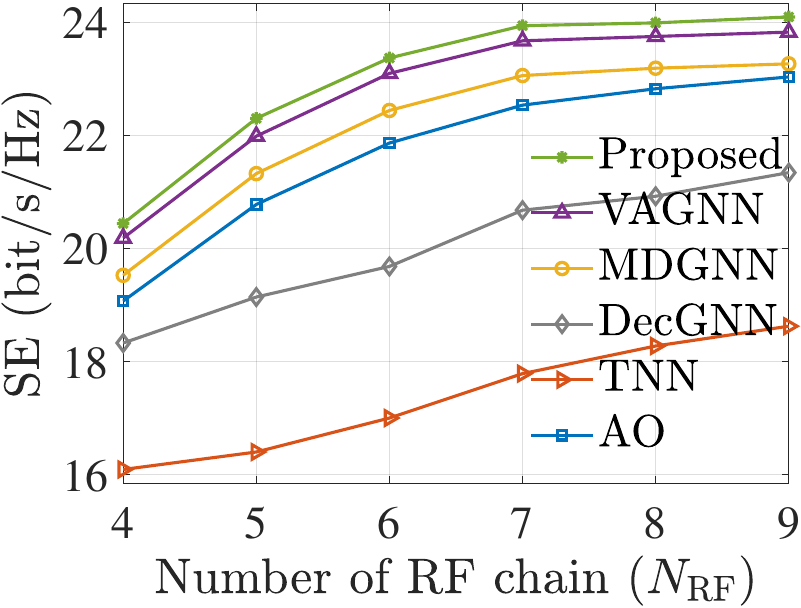}
\label{SE_NRF}
}
\hspace{0.01\linewidth}
\subfloat[SE versus SNR (CDL-D).]{
\includegraphics[width=0.45\linewidth]{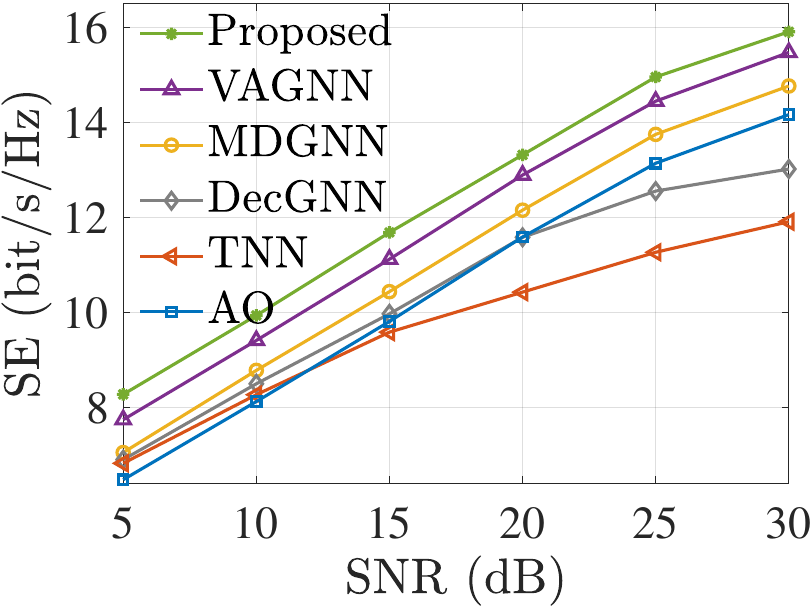}
\label{SE_SNR_CDL}
}
%\vspace{-0.1cm}
\caption{Performance comparison under different parameters and channel models.}
\label{Learning_performance}
\end{figure}

Fig.~\ref{Learning_performance} shows the achievable SE under different system parameters and channel models. As shown in~\figsubref{Learning_performance}{SE_SNR}, the SE increases monotonically with SNR, as expected. As shown in~\figsubref{Learning_performance}{SE_MN}, the SE also increases with the number of antenna elements due to the larger spatial degrees of freedom. The impact of the RF-chain configuration is depicted in~\figsubref{Learning_performance}{SE_NRF}, which shows that increasing $N_{\mathrm{RF}}$ improves all schemes but the gain gradually decreases as more RF chains are added.
In all cases, the proposed method consistently outperforms the baseline schemes. Specifically, it improves the SE by 5.1\%--22.8\% over MDGNN, 11.7\%--28.8\% over DecGNN, and 22.1\%--50.6\% over TNN. The superior performance of the proposed method and VAGNN over MDGNN is mainly attributed to the proposed architecture in Fig.~\ref{MB_DNN}. Moreover, the proposed method outperforms VAGNN due to the gradient-based information aggregation mechanism. TNN performs worse than other learning-based methods because it does not exploit the PE property of the optimal beamforming policy.\footnote{We have also evaluated the fully digital weighted minimum mean squared error (WMMSE) scheme as an upper-bound reference. On average, WMMSE achieves 48.17\% higher SE than the proposed method, which quantifies the performance loss caused by limited RF chains and amplitude-only holographic~beamforming.}

To assess the impact of channel statistics, we also evaluate the schemes under the 3GPP Clustered Delay Line (CDL)-D channel model. As shown in~\figsubref{Learning_performance}{SE_SNR_CDL}, the proposed method maintains its performance advantage under this more realistic propagation model.

%\vspace{-0.2cm}

%\vspace{-0.4cm}
\subsection{Dimension Generalizability}
\begin{figure}[htbp]
\centering
\includegraphics[width=0.41\textwidth]{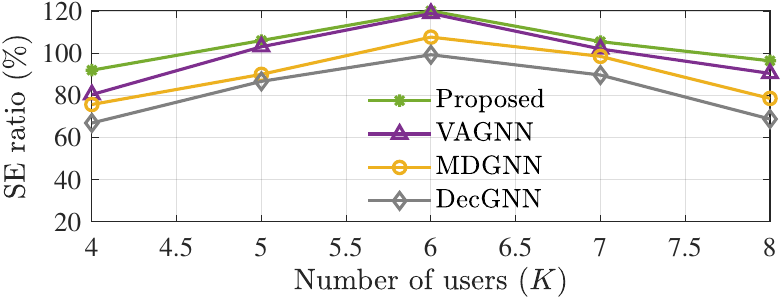}
\caption{Generalization performance versus $K$. }  \label{Generalizability_K}
%\vspace{-0.2cm}
\end{figure}

\begin{figure}[htbp]
\centering
\includegraphics[width=0.41\textwidth]{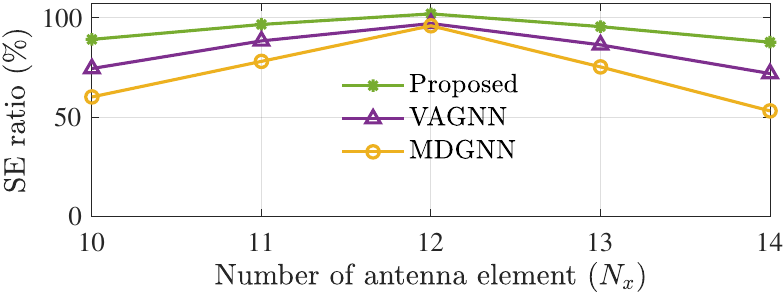}
\caption{Generalization performance versus $N_t$. }  \label{Generalizability_Nt}
%\vspace{-0.4cm}
\end{figure}

Figs.~\ref{Generalizability_K} and~\ref{Generalizability_Nt} evaluate size generalization with respect to the numbers of users and BS antennas, respectively.\footnote{
TNN is excluded from Figs.~\ref{Generalizability_K} and~\ref{Generalizability_Nt}, and DecGNN is excluded from Fig.~\ref{Generalizability_Nt}, since they do not support generalization along the corresponding~dimensions.} For user-dimensional generalization, all methods are trained with $N_{\mathrm{RF}}=10$ and $K=6$, and tested with $K$ ranging from $4$ to $8$. The performance metric is the ratio between the SE achieved by each learning-based method and that achieved by the AO baseline. As shown in Fig.~\ref{Generalizability_K}, the proposed method achieves the highest SE ratio, demonstrating better generalization across different numbers of users than the baseline methods. Compared with MDGNN and DecGNN, it improves user-dimensional generalization performance by 11.6\%--22.7\% and 17.7\%--40.2\%, respectively.

The generalization to the number of antennas is more challenging when the model is trained with a fixed antenna configuration, because the corresponding phase-pattern matrix $\mathbf{M}_{\mathrm p}$ is also fixed. To improve generalizability, we train the models with randomly generated phase-pattern matrices. During training, $N_x$ is sampled from a discretized exponential distribution with a mean of $10$, and each entry phase of $\mathbf{M}_{\mathrm p}$ is independently sampled from a uniform distribution $\mathcal{U}[-\pi,\pi]$. This setup ensures that approximately $80\%$ of the generated samples have $N_x\leq12$, and each sample contains both a random channel matrix $\mathbf{H}$ and a random phase-pattern matrix $\mathbf{M}_{\mathrm p}$. During inference, the trained models are evaluated using the fixed $\mathbf{M}_{\mathrm p}$ corresponding to each tested antenna configuration. 
Fig.~\ref{Generalizability_Nt} shows that this training strategy enables the proposed method to achieve
the best generalization. In particular, it preserves more than 85\% of the AO baseline's performance for both larger ($196$ antennas, i.e., $N_x=N_y=14$) and smaller ($100$ antennas, i.e., $N_x=N_y=10$) antenna~configurations. It improves the generalization performance by 6.2\%--65.1\% over~MDGNN.

%\vspace{-0.1cm}
\subsection{Inference and Training Complexity}
Table~\ref{Complexity} compares the inference and training complexity of the considered methods. Training complexity is evaluated in terms of sample, time, and space requirements, all measured as the minimum training requirement for each learning-based method to reach the SE of the AO baseline.  
 
As shown in Table~\ref{Complexity}, all learning-based methods exhibit significantly lower inference latency than the AO algorithm. For training complexity, the proposed method achieves the lowest sample and space complexity among the learning-based approaches. Its training and inference latency are slightly higher than those of VAGNN, owing to the additional gradient-based operations incorporated into the update equations. MDGNN incurs higher training complexity than the proposed method and VAGNN, since extending the input and output to higher-dimensional representations results in a larger model size and increased number of trainable parameters. Since DecGNN and TNN do not achieve the same SE as the AO baseline, their training complexities are not reported.

%\vspace{-0.3cm}
\begin{table}[htbp]	
\captionsetup{font=small}
\centering 
\small
\caption{Comparison of Inference and Training Complexity}	
\begin{tabular}{c|c|c|c|c}
\hline\hline
\multirow{2}{*}{Name} & \multirow{2}{*}{Inference latency} & \multicolumn{3}{c}{Training complexity} \\
\cline{3-5}
                 &           & Samples       & Time           & Space \\              
\cline{1-5}
\textbf{Proposed} & 20.36 ms   & 25 K      & 3.86 h        & 0.999 M  \\ 
\cline{1-5}
\textbf{VAGNN}    & 8.90 ms    & 50 K      & 2.97 h        & 1.677 M  \\ 
\cline{1-5}
\textbf{DecGNN}   & 2.52 s     &--         &--             & --   \\ 
\cline{1-5}
\textbf{MDGNN}    & 9.71 ms    & 75 K      & 88.05 h       & 13.77 M   \\ 
\cline{1-5}
\textbf{TNN}      & 10.12 ms   & --        & --            & --   \\ 
\cline{1-5}
\textbf{AO}       & 59.23 s    &--         &--             &--  \\ 
\hline\hline
\end{tabular}
\begin{minipage}{0.95\linewidth}
\quad\,\,\footnotesize \textit{Note:} ``K'' and ``M'' denote thousand and million, respectively.
\end{minipage}
\label{Complexity}
%\vspace{-0.2cm}
\end{table}

\section{Conclusions}
This paper proposed a learning-based method for optimizing the HMIMO beamforming. A novel network architecture integrating a GGNN with two cascaded projection modules was developed. The GGNN exploits the joint PEPI property to learn the equivalent and the holographic beamformers, while the two projection modules recover the digital beamformer from the learned equivalent beamformer and enforce the transmit power constraint. Simulation results demonstrate that the proposed method achieves higher SE with lower inference latency than the AO algorithm and exhibits better generalization than other learning-based~baselines.

\bibliographystyle{IEEEtran}
\bibliography{main}
\end{document}